\documentclass[12pt]{article}

\usepackage{scicite}
\usepackage{amsmath}
\usepackage{times}
\usepackage{graphicx}
\usepackage{caption}
\usepackage{color}

\newenvironment{sciabstract}{%
\begin{quote} \bf}
{\end{quote}}

\title{Building one molecule from a reservoir of two atoms}

\author
{L. R. Liu,$^{123}$ J. D. Hood,$^{13}$ Y. Yu,$^{123}$ J. T. Zhang,$^{123}$     N. R. Hutzler,$^{123\dagger}$ \\
T. Rosenband,$^{2}$ and K.-K. Ni$^{\ast123}$\\
\\
\normalsize{$^{1}$~Department of Chemistry and Chemical Biology, Harvard University,}\\
\normalsize{Cambridge, Massachusetts, 02138, USA}\\
\normalsize{$^{2}$~Department of Physics, Harvard University, Cambridge, Massachusetts, 02138, USA}\\
\normalsize{$^{3}$~Harvard-MIT Center for Ultracold Atoms, Cambridge, Massachusetts, 02138, USA}\\
\\
\normalsize{$^{\dagger}$~Current address:  Division of Physics, Mathematics, and Astronomy,} \\
\normalsize{California Institute of Technology, Pasadena, CA, 91125, USA}\\
\normalsize{$^\ast$To whom correspondence should be addressed;  E-mail: ni@chemistry.harvard.edu}\\
}

\date{\today}

\begin{document} 


\baselineskip24pt


\maketitle


\begin{sciabstract}
Chemical reactions typically proceed via stochastic encounters between reactants. Going beyond this paradigm, we combine exactly two atoms into a single, controlled reaction. 
The experimental apparatus traps two individual laser-cooled atoms (one sodium and one cesium) in separate optical tweezers and then merges them into one optical dipole trap. Subsequently, photo-association forms an excited-state NaCs molecule. 
The discovery of previously unseen resonances near the molecular dissociation threshold and measurement of collision rates are enabled by the tightly trapped ultracold sample of atoms. 
As  laser-cooling and trapping capabilities are extended to more elements, the technique will enable the study of more diverse, and eventually more complex, molecules in an isolated environment, as well as synthesis of designer molecules for qubits.

\end{sciabstract}



Chemical reactions proceed through individual collisions between atoms or molecules. However, when performed in  stochastic ensembles, the individual reaction probabilities are observed as averages. Crossed molecular beams reduce the thermal velocity dispersion to probe elementary reaction processes based on single collision events, illuminating many aspects of reaction dynamics~\cite{Herschbach1987, Lee1987,Henson2012,Perreault2017}. In quantum degenerate gases, cooled to temperatures below $1\,\mu$K, the quantum motional degrees of freedom play a critical role in the reaction~\cite{Ospelkaus2010,Ni2010,marcio2011}. Comparisons of  such experimental reaction rates with theoretical models currently underpin our  understanding of reactions at the most elementary level~\cite{Liu2001, Yang2007, Klein2017}. 

To further improve the specificity and precision of reaction steps~\cite{Ratschbacher2012, Moses2015, Puri2017}, individual particle control is needed, similar to pioneering atom-positioning experiments with scanning tunneling microscopes~\cite{Eigler1990}, but untethered from surfaces. By controlling individual particles via laser cooling and optical trapping, molecules may be constructed atom by atom, while maintaining specific internal and external quantum states.

Herein, we realize chemistry in the minimum number regime, where precisely two atoms are brought together to form one molecule with the aid of a photon. We achieve this by using movable optical tweezers, where  individual atoms of different elements (here Na and Cs) are isolated, cooled,  manipulated, and eventually combined into a single optical tweezer. With exactly two atoms in an optical tweezer, we can observe their collisions. We can also perform single molecule spectroscopy in the gas phase by optically exciting the atom pair on a molecular transition, thereby realizing the chemical reaction $Na+Cs\rightarrow NaCs^{\ast}$. Subsequent imaging of Na and Cs fluorescence distinguishes between four possible experimental outcomes: both, only one, or no atoms are detected in the tweezer, the latter indicating a reaction has occurred. 
We chose NaCs for the demonstration because it possesses a large molecular fixed-frame dipole moment of 4.6 Debye~\cite{Dagdigian1972}, making it a strong candidate for a molecular qubit in a future quantum computing architecture.

\begin{figure}[!t]
    \includegraphics[width=100mm]{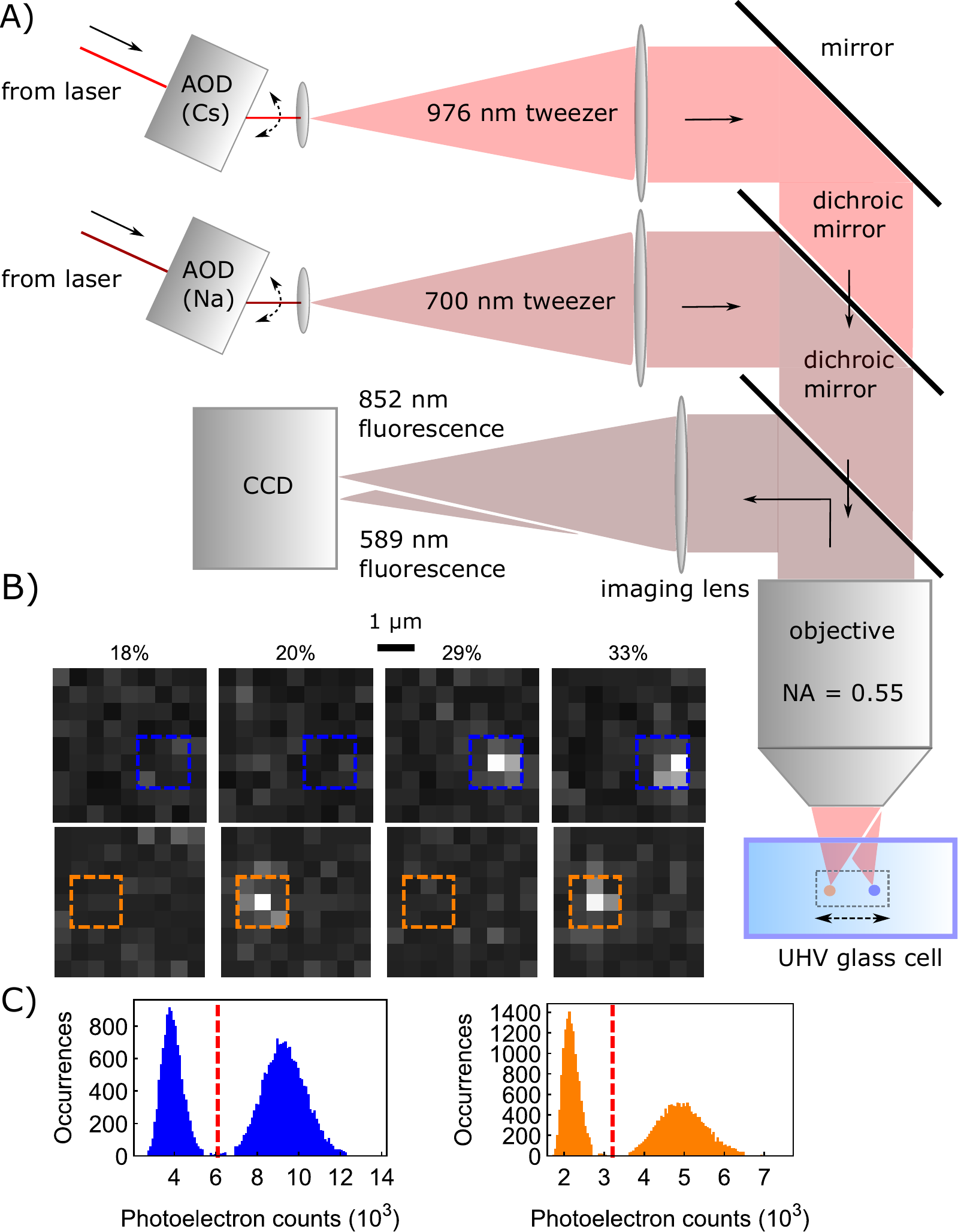}
  \caption{\textbf{Dual-species single atom trapping and imaging.}
 \textbf{A}) Schematic of the setup. Optical tweezer atom trapping beams (700~nm and 976~nm wavelengths) are independently steered by acousto-optic deflectors, expanded by telescopes, and then combined on a dichroic mirror before being focused by the objective into a glass cell. Fluorescence  from trapped Na and Cs atoms is collected through the objective onto the CCD camera. \textbf{B}) Fluorescence images of single Na and Cs atoms. Length scale of $1\,\mu$m is indicated. Cs (top panels) and Na (bottom panels) are imaged sequentially in the same field of view. The four possible cases are shown with their initial loading probabilities: no atoms, a single Na atom, a single Cs atom, both Na and Cs atoms. Dashed blue (Cs) and orange (Na) boxes indicate the region of interest for determining presence of atoms. \textbf{C}) Histogram of Cs (blue) and Na (orange) fluorescence. The bimodal distribution shows clear separation between zero- and one-atom peaks. Red dashed lines indicate the threshold that is used to determine the presence of an atom. }
\label{app}
\end{figure}

\begin{figure*}
  \includegraphics[width=0.7\columnwidth]{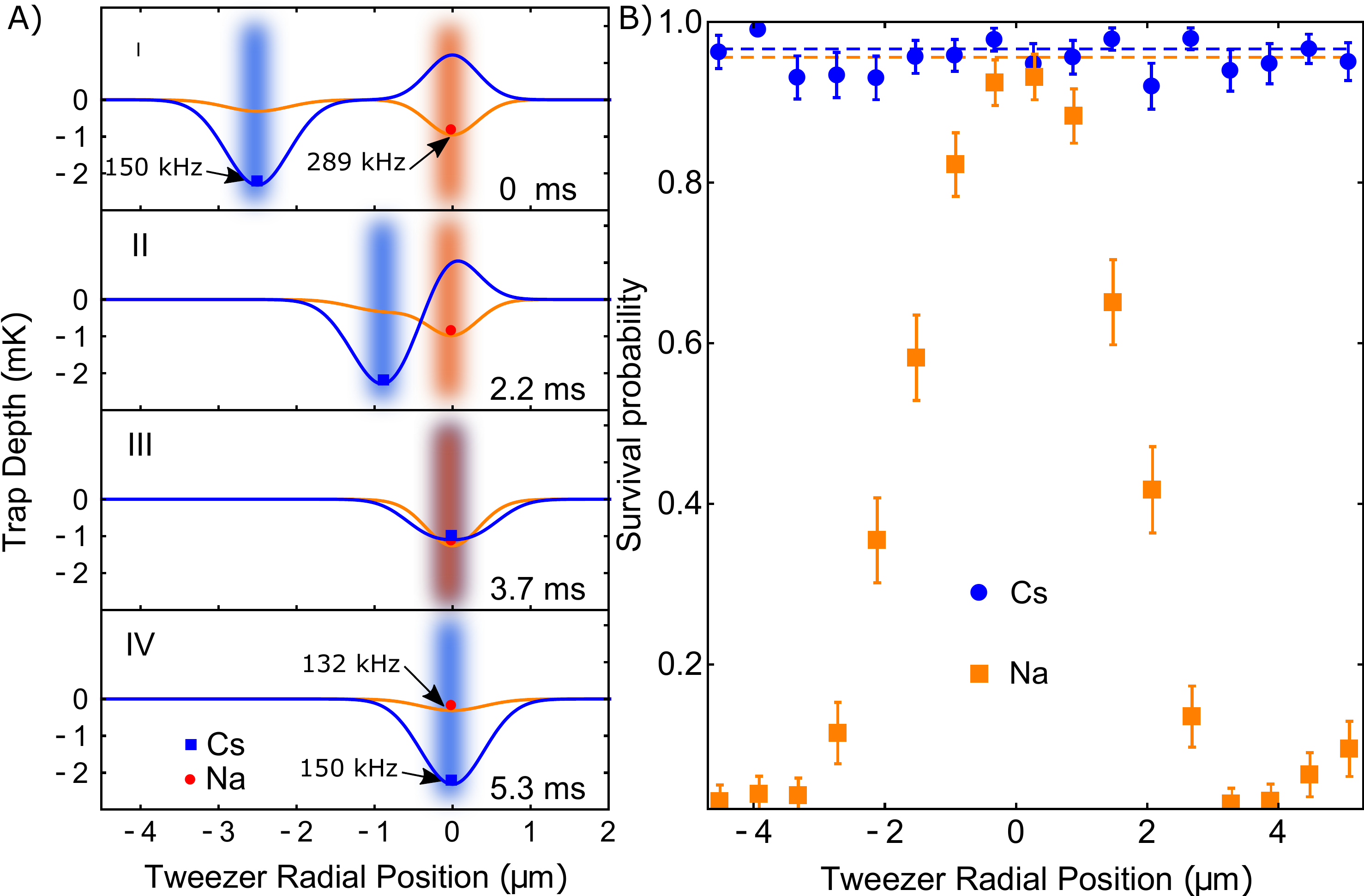}
  \caption{\textbf{Merging single Na and Cs atoms, which are initially separated by 3~$\mu$m, into one tweezer.}
 \textbf{A}) 1-D cuts of the combined, time-varying 700 nm and 976 nm tweezer potentials for both atoms during the merge sequence. Na and Cs are represented by  dots that track the minima of their  potentials (orange for Na and blue for Cs).  Overlaid are graphics of the optical tweezers. Radial trap frequencies are labeled in the first and last panels (axial trap frequencies are roughly 6 times smaller). Panels I-III depict the merging process. In panel IV, the 700 nm tweezer has been extinguished and only the 976 nm tweezer remains.  \textbf{B}) Measured survival probability of Na and Cs after the sequence depicted in (A), followed by separating the tweezers through a reverse sequence to image the atoms. The two atoms are merged into the same tweezer at the survival maximum for Na. Error bars denote the Wilson score interval. The dashed lines represent the survival rates due to imperfect re-imaging.}
\label{merge}
\end{figure*}

We began by preparing laser-cooled Na and Cs atoms at a few hundred $\mu$K in overlapped magneto-optical traps (MOTs) in a  vacuum chamber ($10^{-8}$ Pa).  The MOTs serve as cold atom reservoirs for loading single atoms into tightly focused optical tweezer traps~\cite{Schlosser2001}. After loading, the MOTs are extinguished. A schematic of the apparatus is shown in Fig.~\ref{app}A.  The NA=0.55 microscope objective focuses two different wavelengths of light, 700~nm and 976~nm, to waists of 0.7~$\mu$m and 0.8~$\mu$m radius.  Due to the difference in Na and Cs polarizabilities, the 700~nm wavelength light attracts Na and repels Cs, while 976~nm light attracts Cs  five times more strongly than Na~\cite{Safronova2006}, enabling us to manipulate Na and Cs independently as depicted in Fig.~\ref{merge}A.
A typical trap depth of 1~mK is achieved for 5~mW of tweezer power. 

When tightly confined identical atoms are illuminated with near-atomic-resonant light,  light-assisted pairwise collisions result in either zero or one final atom in the trap~\cite{Schlosser2001,Fuhrmanek2012}. Single atom loading succeeds approximately half of the time~\cite{Sompet2013}.   
However, the large light shifts for Na in a 700~nm wavelength tweezer would normally prevent atom cooling, and consequently, efficient atom loading. We eliminate this light shift for Na by alternating the tweezer and cooling beams at a rate of 3 MHz~\cite{Hutzler2017}. 
Subsequently, Na, followed by Cs, are imaged and polarization gradient cooled to 70~$\mu$K and 10~$\mu$K respectively.
To determine whether an atom is in the optical tweezer, the fluorescence photoelectron counts from each atom in a region of interest  (Fig. \ref{app}B) are compared to a threshold (Fig. \ref{app}C). 
The fluorescence histograms indicate that the cases of zero or one atom can be distinguished with a fidelity better than 99.97~\%.
We find that in 33\% of cases we load a single Na and a single Cs atom  side-by-side. 
In 18\% of cases, no atoms are loaded, and the rest of the time either a single Na or a Cs atom is loaded (Fig. \ref{app}B). 
The experiment, which repeats at 3 Hz, records initial and final fluorescence images to determine survival probabilities for different stages of the molecule formation process. 

Once single atoms have been loaded in separate traps, they need to be  transported to the same location for molecule formation. Optical tweezers have been used to move single atoms while maintaining atomic internal state coherence~\cite{Beugnon2007} and to merge two indistinguishable atoms  by coherent tunneling into one tweezer~\cite{Kaufman2014}. Here we adiabatically transport and merge two different atoms, Na and Cs, into the same tweezer, as depicted in  Fig.~\ref{merge}A  by using optical tweezers at two different wavelengths. The  trap depths are adjusted by changing the  beam intensities, and the positions are steered by applying different radio frequencies to the respective acousto-optic deflectors (AODs) (Fig.~\ref{app}A).

For the merge sequence, the 700~nm tweezer containing Na is kept stationary while the 976~nm tweezer containing Cs is moved to overlap the atoms (Fig.~\ref{merge}A, panel I - III).  Following the merge, the 700~nm tweezer is extinguished adiabatically to leave both atoms in the 976~nm tweezer (Fig.~\ref{merge}A, panel IV). We design this merge trajectory such that i) Cs is deeply confined at all times, and ii) the double-well potential imposed on Na is sufficiently asymmetric to avoid a near-degenerate ground state. This process is time-reversible, which enables us to image the atoms separately and determine survival probability. 

Because the 700~nm tweezer is extinguished for 1~ms after the merge, while the 976 nm tweezer is always active, the Na atom escapes unless the two tweezers are overlapped at the end of the merge sequence, whereas the Cs atom is always trapped.  Fig.~\ref{merge}B shows the result obtained when scanning the endpoint of the 976 nm tweezer trajectory. The height of the Na survival peak at 0~$\mu$m of 94(1)\% is near the re-imaging survival probability of 96\%.

Having demonstrated adiabatic transport and merging of two species into a tight tweezer, we turn to their collisions. Isolated collisions between two atoms do not usually result in molecule formation due to the need to simultaneously conserve momentum and energy. However, the atoms can change their hyperfine states after colliding, and the exothermic hyperfine-spin-changing collisions impart enough kinetic energy ($\approx100$~mK) to the atoms to eject them from the tweezer ($\approx1$~mK depth)~\cite{Ueberholz2000}.

Generally, a given initial trap occupancy can evolve into 4 possible outcomes following an experiment: i) both atoms, ii) no atoms, iii) only Cs, and iv) only Na remain in the trap.  Single-atom images from each repetition allow us to post-select on any of these cases and separate 1- and 2-body processes, giving both  lifetimes from a single dataset (Fig.~\ref{2body}). For example, when Na and Cs are both present (effective pair density of $n_{2} = 2\times10^{12} \, \textrm{cm}^{-3}$~\cite{sm}), and prepared in a mixture of hyperfine spin states, they are both rapidly lost $\tau_{loss} = 8(1)\, \textrm{ms}$, where $\tau_{loss}$ is the $1/e$ time of exponential decay. This yields a loss rate constant $\beta = 5\times10^{-11} \, \textrm{cm}^{3}/\textrm{s}$. In contrast, if the atoms are both optically pumped into the lowest energy hyperfine levels, conservation of energy prevents the change of hyperfine states, and the atom lifetime increases to 0.63(1)~s, similar to the rate of hyperfine-state relaxation for Cs due to off-resonant scattering of the tweezer light~\cite{Cline1994}. When only one atom is present, 1-body loss due to collisions with background gas limits the lifetime to 5 s.

Because of the rapid 2-body loss for mixed hyperfine states, we optically pump each atom into its  lowest energy hyperfine state to maintain a long-lived sample of co-trapped  Na and Cs atoms. 
We then perform photoassociation (PA) of the  atoms to form an excited state molecule, realizing a single instance of the chemical reaction $\textrm{Na}+\textrm{Cs}\rightarrow \textrm{NaCs}^{\ast}$.
When illuminating the atoms with resonant PA light, an electronically excited state molecule may form (Fig.~\ref{PA}A) and then rapidly decay to the ground state. 
The molecule does not scatter imaging light, causing molecule formation to manifest as simultaneous loss of both Na and Cs atoms.  The bottom panel of Fig.~\ref{PA}B shows these loss resonances  as the frequency of the PA light is scanned below the dissociation threshold. 

Our optical tweezer architecture offers a number of advantages for  PA measurements over previous methods with bulk samples~\cite{Jones2006}.  The ability to precisely define the initial reagents eliminates contributions from other reaction processes such as $\textrm{Cs}_2$ formation or 3-body loss.
The combination of the high effective pair density~\cite{sm} $n_{2} = 3\times10^{12} \, \textrm{cm}^{-3}$,
afforded by the tweezer confinement, and high PA light intensity of $3 \, \textrm{kW}/\textrm{cm}^2$  yields fast PA rates.
The high contrast measurements of single-atom loss result in  near-unity molecule detection efficiency and avoid the need for ionization detection~\cite{Jones2006}. 

We scan the 200 MHz frequency-broadened PA light from 30 to 250 GHz below the Cs atomic D2 line  (6S$_{1/2}$ - 6P$_{3/2}$). We take steps of 200 MHz with 100ms pulse duration, and average over approximately 100 repetitions at each data point. An absolute accuracy of 1 GHz is set by the wavemeter.  
During PA, the Cs atom could be promoted into the upper hyperfine level due to off-resonant scattering of the PA beam, which would lead to spin-changing collisional loss. We counteract this effect by simultaneously optically pumping Cs into the lower hyperfine level with a separate beam.

The ability to detect molecule formation via atom loss with high efficiency allows us to probe NaCs* vibrational levels near the dissociation threshold, including resonances that have not been previously observed (Fig.~\ref{PA}).  
According to \textit{ab initio} calculations of NaCs* with spin-orbit coupling~\cite{Korek2007}, five molecular potentials  converge to the Cs~(6P$_{3/2}$) + Na~(3S$_{1/2}$) asymptote (Fig.~\ref{PA}A):  $B^1\Pi_{1}$, $c^3\Sigma^+_{\Omega=0,1}$,  and $b^3\Pi_{\Omega=0,2}$. 
Of these, only the $c^3\Sigma^+_{1}$ levels have previously been observed in the near-threshold regime~\cite{Grochola2011}, and our measurement agrees to within 1~GHz.   
 To identify the vibrational progressions, we fit the LeRoy-Bernstein (LB) dispersion model~\cite{LeRoy1970} to our observed resonances. 
Near threshold, the vibrational quantum number $v'$ ($v'=-1$ is the highest bound state) is related to the  binding energy by 
\begin{equation} \label{eq:LBmodel}
E_{v'} =    -\frac{1}{C_6^{1/2}} \left[ 2 \hbar \left( \frac{2 \pi}{\mu} \right)^{1/2} \frac{\Gamma(7/6)}{\Gamma(2/3)}   (v'-v'_0)\right]^3,
\end{equation} 
where $\mu$ is the reduced mass, and $\hbar$ is the reduced Planck's constant. 
We extract the $C_6$ dispersion coefficients that characterize the $1/r^6$ component of the potentials and $v'_0$, which is an offset between -1 and 0.  

Fitting to the positions of our observed $c^3 \Sigma^+_{1}$ resonances gives $v_0 = -0.79$ and  $C_6 = 8.5(6) \times 10^3 \, \textrm{a.u.}$ (in atomic units),  in agreement with the theoretical value  $C_6 = 7.96 \times 10^3\,\textrm{a.u.}$~\cite{Marinescu1999}.  
From the remaining loss resonances, we identify two additional progressions ($B^1 \Pi_{1}$  and $c^3\Sigma^+_{0}$) with  $C_6 = 1.42(33) \times 10^4\,\textrm{a.u.}$ and  $C_6  = 1.47(26) \times 10^4\,\textrm{a.u.}$  (Fig.~\ref{PA}B).  Both values  are near   the theoretical value of $C_6 = 1.83 \times 10^4\,\textrm{a.u.}$~\cite{Marinescu1999}. Our state labels correspond to the molecular wavefunctions in the near-threshold regime and differ from the labels in Ref.~\cite{Marinescu1999} due to an avoided crossing as  noted in Ref.~\cite{Grochola2010}. Here the assignment of the $c^3\Sigma^+_{1}$ progression is based on previous observation of the same resonances~\cite{Grochola2011}, while  $B^1 \Pi_{1}$  continues a previously observed sequence~\cite{Grochola2010}.  The remaining progression corresponds to $c^3\Sigma^+_{0}$, because this is the only other compatible state. We interpret the photoassociation spectrum as clear evidence for molecule formation, because the resonance peaks appear exclusively as simultaneous loss of Na and Cs, and the resonance frequencies agree with independent measurements.

Our technique can in principle be extended beyond the simple bialkalis demonstrated here, and to produce deeply bound molecules. 
Molecules in a single quantum state could be created by coherent transfer ~\cite{STIRAP, Liu2017} of atoms prepared in the motional ground state~\cite{Monroe1995, Li2012, Kaufman2012,Yu2017}.
Dipolar molecules trapped in a configurable array of optical tweezers~\cite{Barredo2016, Endres2016}  would constitute a new type of qubit for quantum information processing~\cite{Demille2002} and an important resource to explore quantum phases~\cite{Yao2015,Sundar2017}.

\bibliography{refs}

\begin{thebibliography}{10}

\bibitem{Herschbach1987}
D.~R. Herschbach, {\it Angewandte Chemie International Edition in English\/}
  {\bf 26}, 1221 (1987).

\bibitem{Lee1987}
Y.~T. Lee, {\it Angewandte Chemie International Edition in English\/} {\bf 26},
  939 (1987).

\bibitem{Henson2012}
A.~B. Henson, S.~Gersten, Y.~Shagam, J.~Narevicius, E.~Narevicius, {\it
  Science\/} {\bf 338}, 234 (2012).

\bibitem{Perreault2017}
W.~E. Perreault, N.~Mukherjee, R.~N. Zare, {\it Science\/} {\bf 358}, 356
  (2017).

\bibitem{Ospelkaus2010}
S.~Ospelkaus, {\it et~al.\/}, {\it Science\/} {\bf 327}, 853 (2010).

\bibitem{Ni2010}
K.-K. Ni, {\it et~al.\/}, {\it Nature\/} {\bf 464}, 1324 (2010).

\bibitem{marcio2011}
M.~H.~G. {de Miranda}, {\it et~al.\/}, {\it Nature Physics\/} {\bf 7}, 502
  (2011).

\bibitem{Liu2001}
K.~Liu, {\it Annual Review of Physical Chemistry\/} {\bf 52}, 139 (2001).

\bibitem{Yang2007}
X.~Yang, {\it Annual Review of Physical Chemistry\/} {\bf 58}, 433 (2007).
  PMID: 17105413.

\bibitem{Klein2017}
A.~Klein, {\it et~al.\/}, {\it Nature Physics\/} {\bf 13}, 35 EP  (2016).

\bibitem{Ratschbacher2012}
L.~Ratschbacher, C.~Zipkes, C.~Sias, M.~K{\"o}hl, {\it Nature Physics\/} {\bf
  8}, 649 EP  (2012).

\bibitem{Moses2015}
S.~A. Moses, {\it et~al.\/}, {\it Science\/} {\bf 350}, 659 (2015).

\bibitem{Puri2017}
P.~Puri, {\it et~al.\/}, {\it Science\/} {\bf 357}, 1370 (2017).

\bibitem{Eigler1990}
D.~M. Eigler, E.~K. Schweizer, {\it Nature\/} {\bf 344}, 524 EP  (1990).

\bibitem{Dagdigian1972}
P.~J. Dagdigian, L.~Wharton, {\it The Journal of Chemical Physics\/} {\bf 57},
  1487 (1972).

\bibitem{Schlosser2001}
N.~Schlosser, G.~Reymond, I.~Protsenko, P.~Grangier, {\it Nature\/} {\bf 411},
  1024 (2001).

\bibitem{Safronova2006}
M.~S. Safronova, B.~Arora, C.~W. Clark, {\it Phys. Rev. A\/} {\bf 73}, 022505
  (2006).

\bibitem{Fuhrmanek2012}
A.~Fuhrmanek, R.~Bourgain, Y.~R.~P. Sortais, A.~Browaeys, {\it Phys. Rev. A\/}
  {\bf 85}, 062708 (2012).

\bibitem{Sompet2013}
P.~Sompet, A.~V. Carpentier, Y.~H. Fung, M.~McGovern, M.~F. Andersen, {\it
  Phys. Rev. A\/} {\bf 88}, 051401 (2013).

\bibitem{Hutzler2017}
N.~R. Hutzler, L.~R. Liu, Y.~Yu, K.-K. Ni, {\it New Journal of Physics\/} {\bf
  19}, 023007 (2017).

\bibitem{Beugnon2007}
J.~Beugnon, {\it et~al.\/}, {\it Nat Phys\/} {\bf 3}, 696 (2007).

\bibitem{Kaufman2014}
A.~M. Kaufman, {\it et~al.\/}, {\it Science\/} {\bf 345}, 306 (2014).

\bibitem{Ueberholz2000}
B.~Ueberholz, S.~Kuhr, D.~Frese, D.~Meschede, V.~Gomer, {\it Journal of Physics
  B: Atomic, Molecular and Optical Physics\/} {\bf 33}, L135 (2000).

\bibitem{sm}
See supplementary material.

\bibitem{Cline1994}
R.~A. Cline, J.~D. Miller, M.~R. Matthews, D.~J. Heinzen, {\it Opt. Lett.\/}
  {\bf 19}, 207 (1994).

\bibitem{Jones2006}
K.~M. Jones, E.~Tiesinga, P.~D. Lett, P.~S. Julienne, {\it Reviews of Modern
  Physics\/} {\bf 78}, 483 (2006).

\bibitem{Korek2007}
M.~Korek, S.~Bleik, A.~R. Allouche, {\it The Journal of Chemical Physics\/}
  {\bf 126}, 124313 (2007).

\bibitem{Grochola2011}
A.~Grochola, {\it et~al.\/}, {\it Phys. Rev. A\/} {\bf 84}, 012507 (2011).

\bibitem{LeRoy1970}
R.~L. Roy, R.~Bernstein, {\it Journal of Chemical Physics\/} {\bf 52}, 3869
  (1970).

\bibitem{Marinescu1999}
M.~Marinescu, H.~Sadeghpour, {\it Physical Review A\/} {\bf 59}, 390 (1999).

\bibitem{Grochola2010}
A.~Grochola, P.~Kowalczyk, W.~Jastrzebski, {\it Chemical Physics Letters\/}
  {\bf 497}, 22 (2010).

\bibitem{STIRAP}
K.~Bergmann, H.~Theuer, B.~W. Shore, {\it Rev. Mod. Phys.\/} {\bf 70}, 1003
  (1998).

\bibitem{Liu2017}
L.~R. {Liu}, {\it et~al.\/}, {\it ArXiv:1701.03121\/}  (2017).

\bibitem{Monroe1995}
C.~Monroe, {\it et~al.\/}, {\it Phys. Rev. Lett.\/} {\bf 75}, 4011 (1995).

\bibitem{Li2012}
X.~Li, T.~A. Corcovilos, Y.~Wang, D.~S. Weiss, {\it Phys. Rev. Lett.\/} {\bf
  108}, 103001 (2012).

\bibitem{Kaufman2012}
A.~M. Kaufman, B.~J. Lester, C.~A. Regal, {\it Phys. Rev. X\/} {\bf 2}, 041014
  (2012).

\bibitem{Yu2017}
Y.~{Yu}, {\it et~al.\/}, {\it ArXiv:1708.03296\/}  (2017).

\bibitem{Barredo2016}
D.~Barredo, S.~de~L{\'e}s{\'e}leuc, V.~Lienhard, T.~Lahaye, A.~Browaeys, {\it
  Science\/} {\bf 354}, 1021 (2016).

\bibitem{Endres2016}
M.~Endres, {\it et~al.\/}, {\it Science\/} {\bf 354}, 1024 (2016).

\bibitem{Demille2002}
D.~DeMille, {\it Phys. Rev. Lett.\/} {\bf 88}, 067901 (2002).

\bibitem{Yao2015}
N.~Y. Yao, M.~P. Zaletel, D.~M. Stamper-Kurn, A.~Vishwanath, {\it arXiv
  preprint arXiv:1510.06403\/}  (2015).

\bibitem{Sundar2017}
B.~Sundar, B.~Gadway, K.~R. Hazzard, {\it arXiv preprint arXiv:1708.02112\/}
  (2017).

\bibitem{Tuchendler2008}
C.~Tuchendler, A.~M. Lance, A.~Browaeys, Y.~R. Sortais, P.~Grangier, {\it
  Physical Review A\/} {\bf 78}, 033425 (2008).

\end{thebibliography}

\bibliographystyle{Science}

\section*{Acknowledgments} 

We thank R. Gonz\'alez-F\'erez, P. Julienne, D. DeMille, and C. Regal for discussions. K.-K. N. thanks D. S. Jin for encouragement to pursue the research  presented here. Funding: This work is supported by the Arnold and Mabel Beckman Foundation, as well as the AFOSR Young Investigator Program, the NSF through the Harvard-MIT CUA, and the Alfred P. Sloan Foundation. Author contributions: L. R. L., J. D. H., Y. Y., J. T. Z., N. R. H., T. R., K.-K. N. performed the experiment. L. R. L., J. D. H., T. R., K.-K. N. analyzed the data and wrote the manuscript. Competing interests: None. Data and materials availability: All data are supplied in the paper and supplementary material.

\section*{Supplementary materials}
Materials and Methods\\
Supplementary Text\\
Fig. S1\\
Table S1\\

\begin{figure*}
  \includegraphics[width=\columnwidth]{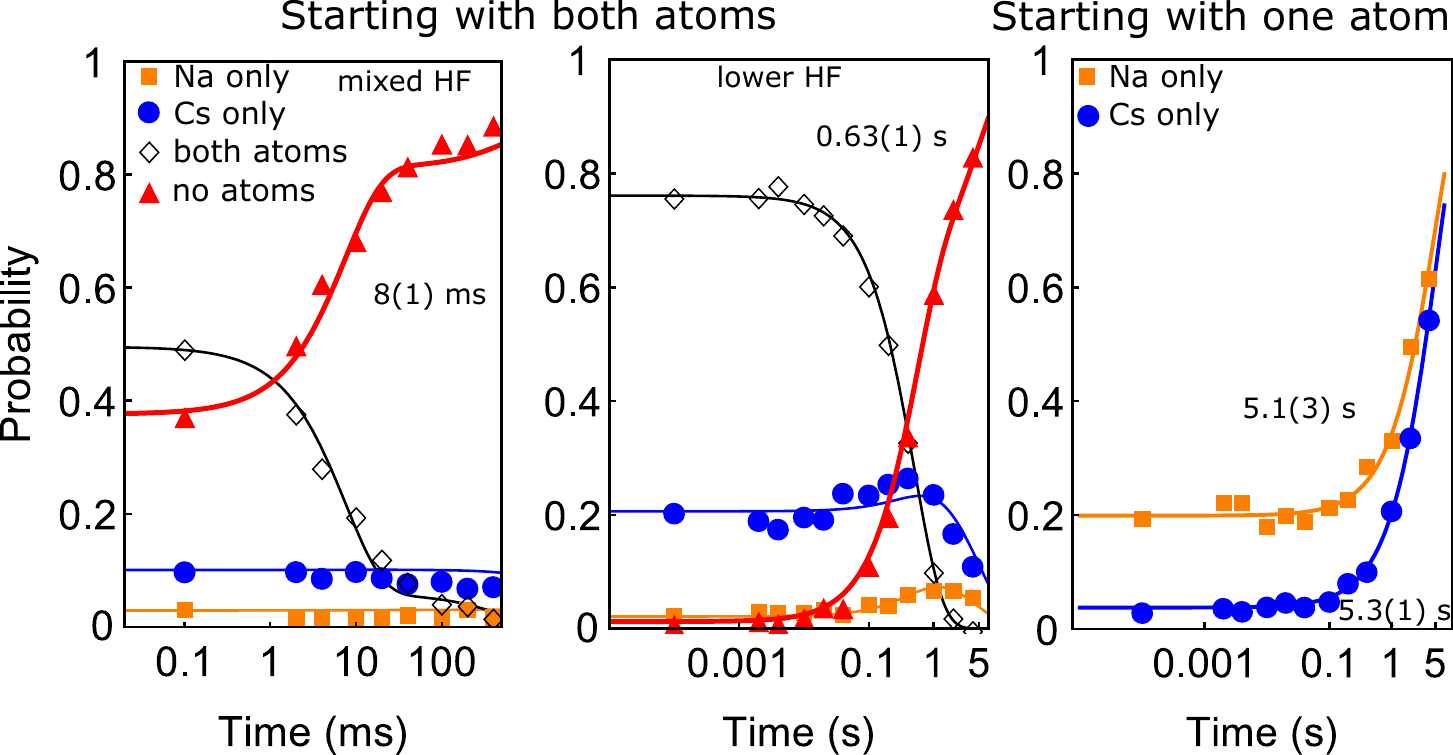}
  \caption{
\textbf{Collisions of Na and Cs.}
The hold time in the merged trap is varied to measure the evolution of trap occupancy due to various collision mechanisms.  Post-selection on initial and final trap occupancies allows us to distinguish 1- and 2-body processes. The fastest timescales are indicated next to the thick fitted curves. The fits are explained in the supplementary material. \textbf{Left}: For both atoms in a mixture of hyperfine states, the loss is dominated by rapid 2-body hyperfine-state-changing collision induced loss. \textbf{Center}: For both atoms in their lowest hyperfine states, the loss is explained by 2-body hyperfine state changing collisions that follow off-resonant scattering of trap light. In these two panels, different markers denote the final trap occupancy. \textbf{Right}: One-body loss gives background gas limited lifetime of about 5~s for both atoms. Here, we post-select on empty final tweezers  and markers denote initial trap occupancy.}

\label{2body}
\end{figure*}

\begin{figure*}
  \includegraphics[width=\columnwidth]{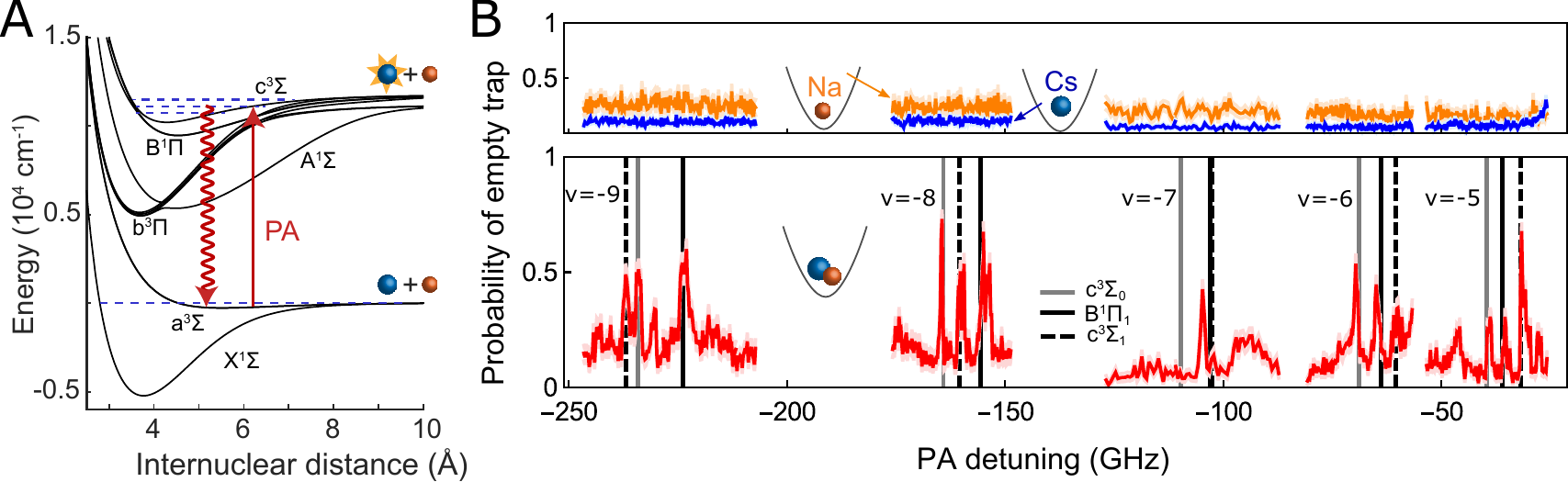}
  \caption{\textbf{Photoassociation Spectroscopy of NaCs$^*$.}  \textbf{A})  NaCs molecular potentials as a function of internuclear distance \cite{Korek2007}.   Photoassociation (PA) light excites the ground state atoms to vibrational levels of the NaCs$^*$ excited molecular potentials, from which they mostly decay to vibrationally excited electronic ground state molecules (squiggly line). The long range asymptotes of the excited state potentials (dominated by van der Waals interactions in the heteronuclear molecules) correspond to one of two cases:  ground state Na colliding with excited Cs in either the lower energy $6P_{1/2}$ (D1 line) or higher energy $6P_{3/2}$ state (D2 line). 
\textbf{B}) The probability of single Na (orange), Cs (blue), and joint Na+Cs (red) atoms evolving to the ``no atoms" detection channel, as the PA light is detuned from the Cs D2 line dissociation threshold at 351730 GHz. When both atoms are initially loaded into the tweezer (lower panel), 2-body loss resonances appear due to molecule formation. As a validation of our method, we check that
no loss resonances are observed  when only one atom is present (upper panel).
The positions of the loss resonances are fitted with the LB dispersion model in eq.~\ref{eq:LBmodel} to identify three different potentials and fit the respective $C_6$ dispersion coefficients. 
The expected resonance positions based on these fits are marked by vertical lines as indicated in the legend. Except for at v=-7, the RMS deviation of the fitted dispersion curve from the measured frequencies are 0.3, 0.6, and 0.8 GHz for the c$^3\Sigma_1$, c$^3\Sigma_0$, and B$^1\Pi_1$ states, respectively. At v=-7, a crossing of molecular energy levels causes the measured spectrum to deviate from the prediction based on eq.~\ref{eq:LBmodel}. Unassigned lines in the spectrum are likely due to rotational and hyperfine structure and pre-dissociating potentials.}
\label{PA}
\end{figure*}


\newpage
\setcounter{equation}{0}    
\setcounter{page}{1} 
\setcounter{figure}{0}    
\renewcommand{\thepage}{S\arabic{page}}  
\renewcommand{\thesection}{S\arabic{section}}   
\renewcommand{\thetable}{S\arabic{table}}   
\renewcommand{\thefigure}{S\arabic{figure}}
\renewcommand{\theequation}{S\arabic{equation}}

\begin{centering} \textbf{MATERIALS AND METHODS}\\
\end{centering}

\begin{centering} \textbf{Experimental apparatus}\\
\end{centering}

All experiments are performed in an epoxy-bonded quartz vacuum chamber maintained at  $<10^{-8}$ Pa by a getter/ion pump.
Six alkali dispensers for Cs and Na are suspended inside the chamber, 5 inches from the MOT region. We maintain a constant current of 2A and 4 A through two of them to maintain sufficient Cs and Na vapor pressures the experiments presented.
 
The 700 nm optical tweezer is derived from a cavity-locked Titanium sapphire laser. The 976 nm tweezer is derived from a free-running distributed Bragg reflector (DBR) laser.

The light for the Na MOT is derived from a frequency doubled Raman fiber amplifier that is seeded by a 1178 nm external cavity diode laser (ECDL). The cooling and the repumping frequencies are generated from the same laser by sending it through a 1.7 GHz acousto optical modulator.
Optical pumping for Na is provided by another 1178 nm ECDL that is frequency doubled with a PPLN waveguide. Optical pumping on the D1 instead of D2 line is necessary to avoid unwanted off-resonant cycling transitions in the Na D2 line due to the small excited-state hyperfine splitting. All lasers for Na are locked via saturated absorption spectroscopy to a vapor cell.

The light for the Cs MOT is derived from two 852 nm DBR lasers. The first is locked to the Cs D2 line via saturated absorption spectroscopy, while the second is referenced to the first with a phase lock, providing repumping and cooling frequencies respectively. 
Optical pumping for Cs is provided by the same light.

At the vacuum chamber, 2 mW of Cs MOT and 5 mW of Na MOT light are expanded to 6 mm beam diameter and combined before being directed into the chamber in a 6-beam configuration. The gradient field for both MOTs, which are formed simultaneously, is 9 G/cm. A 0.55 NA achromatic objective points at the chamber from between two of the horizontal MOT beams. A custom dichroic separates resonant fluorescence from the far detuned tweezers for both species simultaneously. 

We image atoms onto an EMCCD with a magnification factor of 16. The total efficiency from atomic fluorescence to photoelectron counts is about 4$\%$.
To determine the presence of atoms, we image them with resonant light. The signal is  $\sim 10^{4}$ counts/s, at imaging times of 1 ms and 10 ms for Na and Cs respectively.

\begin{centering} \textbf{Na and Cs effective pair density}\\
\end{centering}
 The ``effective pair density" $n_{2}$ is defined as the probability of finding a single Na and Cs atom per unit volume (eq.~\ref{eq:pairdensity})
\begin{equation} 
\label{eq:pairdensity}
n_{2}=\int\limits_{-\infty}^{\infty}\int\limits_{-\infty}^{\infty}\int\limits_{-\infty}^{\infty} n_{Cs}(x,y,z) n_{Na}(x,y,z)\, \textrm{d}x\, \textrm{d}y\, \textrm{d}z
\end{equation}
To get the individual atomic density distributions $n_{Na}(x,y,z)$ and $n_{Cs}(x,y,z)$, we assumed the atoms occupy a thermal ensemble in a 3-dimensional harmonic oscillator potential with trap frequencies (132, 123, 24) kHz for Na and (150, 140, 28) kHz for Cs, as measured by parametric heating. 
The temperature during the collision measurements was measured  by a release and recapture technique~\cite{Tuchendler2008} and found to be 90 $\mu\textrm{K}$ and 42 $\mu\textrm{K}$ for Cs and Na, respectively, giving $n_{2} = 2\times10^{12} \, \textrm{cm}^{-3}$. For the PA measurements the Cs temperature was 28 $\mu\textrm{K}$, giving $n_{2} = 3\times10^{12} \, \textrm{cm}^{-3}$.

\begin{centering} \textbf{Na and Cs 1- and 2-body collisions}\\
\end{centering}

\begin{figure*}
  \includegraphics[width=\columnwidth]{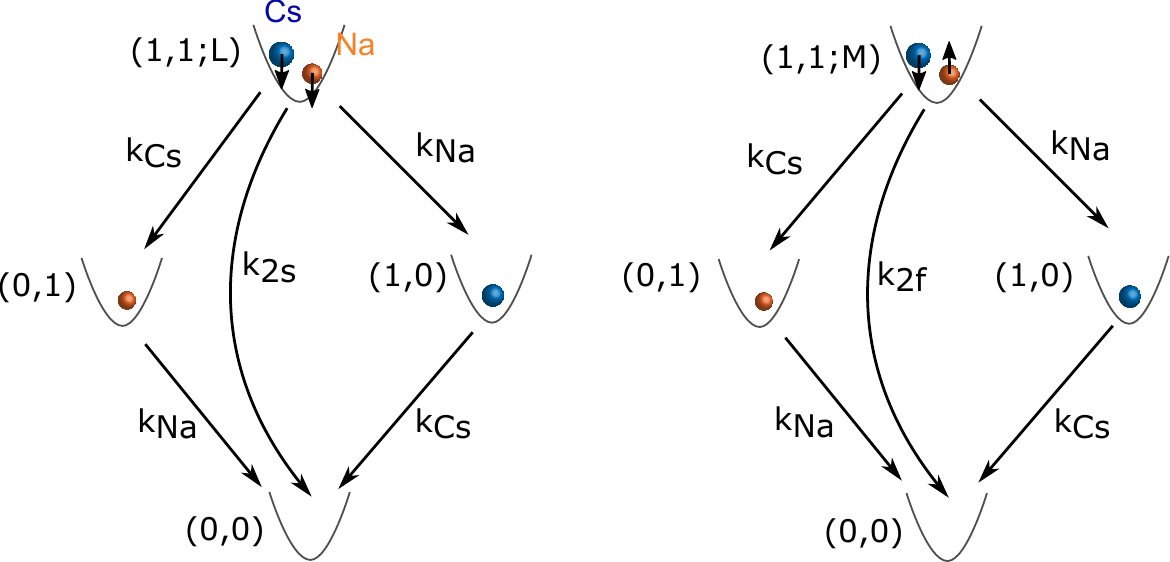}
\caption{
\textbf{Model for 2-body collisions of Na and Cs.} 
Four possible tweezer occupation states exist: (1,1) both Cs and Na; (0,1) only Na; (1,0) only Cs; (0,0) empty. Transitions between states are depicted by arrows with associated rates: 1-body Cs loss $k_{Cs}$, 1-body Na loss $k_{Na}$, slow 2-body loss $k_{2s}$, fast 2-body loss $k_{2f}$. Single atom images allow us to directly detect transitions between any two of these states, thereby determining the rates $k$. (1,1) is further split into two components: L, where both Na and Cs are in their lowest hyperfine states; and M, any other combination of hyperfine states. }
\label{rateEq}
\end{figure*}

To obtain the fits in Fig.~\ref{2body}, we use the model depicted in Fig.~\ref{rateEq}. This yields the system of differential equations eq.~\ref{eq:2body} for the time dependence of each tweezer occupation state. 
The boundary conditions are the initial populations of each state (which can be read off directly from the data) and the fact that all population should end up in (0,0) at long times.

Single atom images and post-selection allow us to isolate individual branches of Fig.~\ref{rateEq}. The 1-body processes ((1,0) to (0,0) and (0,1) to (0,0)) feature only a single exponential decay and are fitted first to obtain $1/k_{Cs}=5.3(1)\,\textrm{s}$ and $1/k_{Na}=5.1(3)\,\textrm{s}$, (Fig.~\ref{2body}, Right). These rates are then fixed and the losses out of (1,1;L) are fitted to obtain $1/k_{2s}=0.63(1)\,\textrm{s}$ (Fig.~\ref{2body}, Center). Finally, this rate is fixed as well and the losses out of (1,1;M) are fitted to obtain $1/k_{2f}=8(1)\,\textrm{ms}$ (Fig.~\ref{2body}, Left).

\begin{equation} 
\label{eq:2body}
\frac{d}{dt}
\begin{bmatrix}
P_{00}(t)\\
P_{01}(t)\\
P_{10}(t)\\
P_{11;L}(t)\\
P_{11;M}(t)
\end{bmatrix}
=
\begin{bmatrix}
    0      & k_{Na} &k_{Cs} &k_{2s} &k_{2f} \\
    0      &- k_{Na} &0 &k_{Cs} &k_{Cs}  \\
    0     &0 &-k_{Cs} &k_{Na} &k_{Na} \\ 
    0    &0 &0 &-k_{2s}-k_{Cs}-k_{Na} &0 \\
    0    &0 &0 &0 &-k_{2f}-k_{Cs}-k_{Na} \\
\end{bmatrix}
\begin{bmatrix}
P_{00}(t)\\
P_{01}(t)\\
P_{10}(t)\\
P_{11;L}(t)\\
P_{11;M}(t)
\end{bmatrix}
\end{equation}

 For the measurements of 2-body collisions, the Cs and Na temperatures are measured to be 90 $\mu\textrm{K}$ and 42 $\mu\textrm{K}$ respectively, giving $n_{2} = 2.3\times10^{12} \, \textrm{cm}^{-3}$. This yields a loss rate constant $\beta = 5\times10^{-11} \, \textrm{cm}^{3}/\textrm{s}$.

\begin{centering} \textbf{NaCs* Photoassociation spectroscopy}\\
\end{centering}

The NaCs photoassociation spectroscopy data presented in Fig.~\ref{PA} are tabulated in \text{Table\_S1.txt}.
The data are organized in columns. The first column is the PA laser detuning in GHz. The next columns are probability, followed by associated error bar, for the outcomes  (Cs,Na) = (1,1) to (0,0), (1,1) to (0,1), (1,1) to (1,0), (1,1) to (1,1), (0,1) to (0,0), and (1,0) to (0,0).

\clearpage


\end{document}